\documentclass[useAMS,usenatbib]{mn2e}

\usepackage{epsfig}

\usepackage{times}

\title[GRB Variability/Peak Luminosity with BATSE GRBs]
{Testing the GRB Variability/Peak Luminosity
correlation using the pseudo-redshifts of a large sample of BATSE GRBs}
\author[C. Guidorzi]{Cristiano Guidorzi$^{1}$\thanks{E-mail:
crg@astro.livjm.ac.uk}\\
$^{1}$Astrophysics Research Institute, Liverpool John Moores University,
Twelve Quays House, Birkenhead, CH41 1LD, UK}
\begin{document}

\date{}

\pagerange{\pageref{firstpage}--\pageref{lastpage}} \pubyear{2002}

\maketitle

\label{firstpage}

\begin{abstract}
We test the correlation found by \citet{Reichart01} between time variability
and peak luminosity of Gamma-Ray Bursts (GRBs). Recently \citet{Guidorzi05}
found that this still holds for a sample of 32 GRBs with spectroscopic redshift,
although with a larger scatter than that originally found by \citet{Reichart01}.
However \citet{Guidorzi05} also found that a power law does not provide a good description of that.
We report on the same test performed on a sample of 551 BATSE GRBs with a significant
measure of variability assuming the pseudo-redshifts derived by \citet{Band04} (1186 GRBs)
through the anticorrelation between spectral lag and peak luminosity.
We still find a correlation between variability as defined
by \citet{Reichart01} and peak luminosity with higher significance.
However, this subsample of BATSE GRBs show a higher scatter around the best-fitting power law
than that found by \citet{Reichart01} in the variability/peak luminosity space.
This is in agreement with the result found by \citet{Guidorzi05}
on a sample of 32 GRBs with measured redshift. These results confirm that a power law
does not provide a satisfactory description for all the GRBs, in contrast with the original
findings by \citet{Reichart01}.

\end{abstract}


\section{Introduction}
Years after the discovery of the first X-ray \citep{Costa97} and optical
\citep{Paradijs97} afterglow counterparts to Gamma-Ray Bursts (GRBs) and the
determination of the first redshift \citep{Metzger97}, it has been possible
to discover some correlations between burst-frame properties derived
on a sample of some tens of GRBs with measured spectroscopic redshift.
Some of them involve spectral and temporal properties:
e.g., the anticorrelation between peak luminosity and spectral lag
discovered by \citet{Norris00}, and the correlation between
peak luminosity and temporal variability found by \citet{Fenimore00}
and \citet{Reichart01} (hereafter FRR00 and R01, respectively).
Other correlations concern the spectrum and the energetics 
of the GRBs, like the \citet{Amati02} correlation between the peak energy
$E_{\rm p}$ of the burst-frame $E\,F(E)$ spectrum and the total isotropic-equivalent
released energy, or the \citet{Ghirlanda04} one between $E_{\rm p}$ and
the collimation-corrected released energy for those GRBs for which it
was possible to estimate the beaming angle. Similarly to the Amati relationship,
\citet{Yonetoku04} found that $E_{\rm p}$ also correlates
with the peak luminosity $L$.
Some of these relationships have been used to estimate the redshifts
of large samples of GRBs.
In particular, \citet{Band04} (hereafter BNB04) used the anticorrelation
between peak luminosity and spectral lag to estimate the redshift of
1186 BATSE GRBs.

Recently, the debate on the Amati and Ghirlanda correlations saw a couple
of papers, \citet{Nakar05} and \citet{BP05}, according to which at least a considerable fraction
of the overall catalogue of BATSE GRBs cannot be consistent with those correlations.
On the other side, three other papers suggest that the majority of the BATSE GRBs are
consistent with the Amati and Ghirlanda correlations: \citet{Ghirlanda05},
\citet{Bosnjak05} and \citet{Pizzichini05}. These authors, starting from different
assumptions concerning the redshifts of the BATSE GRBs, independently show that the
Amati and Ghirlanda correlations are confirmed. See the review by \citet{Amati05}
for a detailed discussion of the debate.


Concerning the correlation originally found by R01 between variability
and peak luminosity, also confirmed for X-Ray Flashes (XRFs; see \citet{Heise01})
by \citet{Reichart03}, recently \citet{Guidorzi05} (hereafter G05)
have confirmed it through a larger sample of 32 GRBs with measured redshift,
although with several differences from R01.
In fact, they find a much larger scatter than that found by R01 with
some notable outlier. Furthermore, G05 found that
the power-law description originally obtained by \citet{Reichart01} for
a sample of 13 GRBs is no more satisfactory for an enlarged sample of
32 GRBs with measured redshift.

In this paper we test the variability vs. peak luminosity correlation
similarly to G05, this time using a large sample of BATSE
GRBs assuming the pseudo-redshifts derived from BNB04.
In particular, we used a different approach from \citet{Ghirlanda05}:
since BNB04 do not provide confidence intervals on both redshift $z$
and peak luminosity $L$, following BNB04 we assumed the validity of the
\citet{Norris00} anticorrelation between peak luminosity and burst-frame
spectral lag and for each BNB04 GRB we derived confidence intervals
on both $z$ and $L$ and made sure that they were consistent with the
BNB04 catalogue.
Eventually we compared the results with those by G05.

The idea of using a large sample of BATSE GRBs with unknown redshift
to test the variability/peak luminosity correlation is not new:
\citet{Schaefer01} used 112 BATSE GRBs to test the correlation between
spectral lag and variability, as defined by FRR00,
as both have been shown to be correlated with peak luminosity.
The motivation of the investigation reported in this paper
is based on several aspects. Firstly,
it must be pointed out that \citet{Schaefer01} made use
of different definitions of both variability and peak luminosity,
very similar to those given by FRR00.
Although the two definitions of variability given by R01 and FRR00
appear to be correlated (see fig.~3 in FRR00), the relation
between the two does not seem to be direct proportionality.
In fact, \citet{Schaefer01} found a best-fitting power-law index
between $L$ and $V$ of $m_{\rm S01}=2.5\pm1.0$ to be compared with
that found by R01, $m_{\rm R01}=3.3^{+1.1}_{-0.9}$.
Secondly, nowadays the greater number of GRBs with known redshift
than that available at that time, allowed to refine the best-fitting
parameters of the lag/luminosity correlation \citep{Norris02}.
For the first time, here we compare the properties in the variability/peak
luminosity space of a sample of 32
GRBs with measured redshift studied by G05 with those of a large
BATSE GRBs sample, using the definitions by R01 for variability
and peak luminosity.

In Section~\ref{s:obs} we discuss how we derived the samples of GRBs.
In Section~\ref{s:var} we describe how we calculated the
variability. In Section~\ref{s:results} we present our results
and in Section~\ref{s:disc} we discuss them.

\section[]{The GRB sample}
\label{s:obs}
%
From the catalogue published by BNB04 we first selected 1186 GRBs,
for which all the information derived by BNB04 is available:
redshift, spectral lag, best-fitting spectral parameters, peak luminosity.
After requiring a significantly positive time lag, this sample
shrank to 866 GRBs. Since BNB04 do not provide uncertainties
on their measures of redshift and peak luminosity, we evaluated
them as follows. We used the photon peak count rates (ph cm$^{-2}$~s$^{-1}$) in the
50--300~keV band measured on a 256-ms time-scale, as reported in the BATSE catalogue
\footnote{http://gammaray.msfc.nasa.gov/batse/grb/catalog/current/tables/flux\_table.txt}.
In order to evaluate the energy peak flux (erg cm$^{-2}$~s$^{-1}$) we need
to know the spectrum. We adopted the same values as BNB04: i.e., 
we made use of the best-fitting spectral parameters
found by \citet{Mallozzi98} who fitted the energy spectrum of the peak
with the Band function \citep{Band93}. For those GRBs for which this
piece of information was not available, like BNB04 we also assumed the average
values found by \citet{Preece00} $\alpha=-0.8\pm0.1$ and $\beta=-2.3\pm0.1$,
while $E_{\rm p}=E_0\,(2+\alpha)$ was taken from BNB04.
From the photon peak flux we calculated the normalisation.
We then evaluated the bolometric energy peak flux $F_{\rm B}$ in the
0--10$^4$~keV band and its uncertainty.

\begin{figure}
\includegraphics[width=8.5cm]{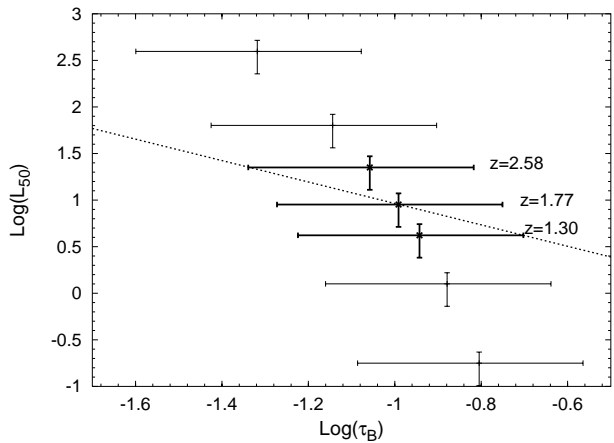}
\caption{Example of how we determined the redshift confidence interval in the
case of trigger \#~228. Dashed line shows the Norris relation. Error bars are
at 2-$\sigma$. The result for this GRB is $z=1.77_{-0.47}^{+0.81}$. BNB04 found
for this GRB $z=1.435$.}
\label{f:vai_0228}
\end{figure}

We calculated the best value and a 2-$\sigma$ confidence interval for the redshift
of each GRB assuming the validity of the anticorrelation between burst-frame
spectral lag $\tau_{\rm B}$ and bolometric peak luminosity $L_{\rm B}$ originally
found by \citet{Norris00} and refined by \citet{Norris02}:
$L_{50} = 21.8\, (\tau_{\rm B}/0.35\, {\rm s})^{-1.15}$,
$L_{\rm B}=L_{50}\times 10^{50}$~erg~s$^{-1}$.
We determined the range for the redshift $z$ for which the point
$(\tau_{\rm B}, L_{\rm B})$ was consistent within 2-$\sigma$ with the Norris
relation (see Fig.~\ref{f:vai_0228}).
We used the following: $\tau_{\rm B}=\tau_0/(1+z)^c$ ($\tau_0$ is the spectral lag
measured in the observer frame; $c=0.6$, that accounts for both cosmological dilation
and narrowing of pulses with higher energy found by \citet{Fenimore95}) and
$L_{\rm B} = 4\pi\,D_{\rm L}^2(z)\,F_{\rm B}$, where $D_{\rm L}(z)$ is the luminosity
distance at redshift $z$. Here we adopted the same cosmology as BNB04:
$H_0 = 70$ km s$^{-1}$ Mpc$^{-1}$, $\Omega_m = 0.3$, and $\Omega_{\Lambda} = 0.7$.
%
%

We point out that the uncertainties on $\tau_{\rm B}$ and
$L_{\rm B}$ have been propagated from those affecting the measures of the lag itself
and of the energy peak flux, respectively (at each fixed $z$).
We studied the scatter distribution of the 2-$\sigma$ lower and upper limits
with respect to the Norris power-law: it turned out that for the 67\% of
the entire sample these limits are scattered more than $0.26$ around the best-fitting
power law. Since \citet{Schaefer01} found a scatter of $0.26$ we conclude
that at least a fraction of 67\% have estimated luminosities that are
consistent with the scatter intrinsic to the lag-luminosity relation.

This is different from the approach followed
by \citet{Ghirlanda05}, who assigned each GRB peak luminosity an uncertainty 
derived from its scatter with respect to the Norris relation (see their eq.~5).
Understandably, our confidence intervals for $z$ are fully consistent with
the values provided by BNB04 for the entire sample of 866 GRBs so far considered.

Like \citet{Ghirlanda05}, we excluded 38 GRBs with $E_{\rm p}\le 40$~keV because
close to the lower boundary of BATSE energy pass-band and therefore likely to be
biased. In addition, one GRB was rejected for its undetermined redshift range,
leaving us with a BNB04 final subset of 827 GRBs.

Eventually, in order to comply with R01 energy band and cosmology, making
use of the same spectral parameters as above, we derived the best peak
luminosity and its 2-$\sigma$ confidence interval in the burst-frame 100--1000~keV
energy band from the best value for $z$ and its 2-$\sigma$ confidence interval,
assuming this time $H_0 = 65$ km s$^{-1}$ Mpc$^{-1}$ like R01 and G05.
Let $\Phi(E)$ be the photon spectrum at peak (ph cm$^{-2}$~s$^{-1}$); $L$ in the
burst-frame 100--1000~keV was then computed according to eq.~\ref{eq:lum}.
\begin{equation}
\displaystyle L \ =\  4 \pi D_L^2(z)\ \int_{100/(1+z)}^{1000/(1+z)} E \Phi(E)\,dE
\label{eq:lum}
\end{equation}
This is formally the same as eq.~9 in R01 and eq.~8 in G05.

\section{Variability Estimation}
\label{s:var}
Variability $V$ was estimated using the public BATSE 64-ms
concatenated light curves
\footnote{ftp://cossc.gsfc.nasa.gov/compton/data/batse/ascii\_data/64ms}.
For each GRB we interpolated the background by fitting with
polynomials of up to 4$^{\rm th}$ degree as prescribed by the BATSE team
\footnote{http://cossc.gsfc.nasa.gov/batse/batseburst/sixtyfour\_ms/bat\_files.revamp\_join}.
This procedure was applied independently to each of the four energy channels:
25--55 keV, 55--110 keV, 110--320 keV, and $>$~320 keV.
Variability was computed for each GRB independently in each channel
according to the definition by R01 (eqs. 4--8 therein).
Following R01, we adopted a smoothing time-scale of $T_f$ with $f=0.45$:
$T_f$ is the shortest cumulative time interval in which a fraction $f$
of the total fluence is collected. Hereafter $V_f$ is the variability
obtained adopting a time-scale of $T_f$.
Uncertainties on $V_f$ have been calculated combining the statistical
uncertainty expressed by eq.~8 of R01 with that due to the error
on redshift $z$.

Eventually, for each GRB we performed a $\chi^2$ test and rejected
all the GRBs showing significantly different variability measures
between different energy channels.
This requirement relies on the definition of $V_f$:
as explained by R01, the definition of variability already
accounts for the narrowing of pulses with energy \citep{Fenimore95}
(pulses' width is proportional to $E^{-\alpha}$, $\alpha=0.4$)
and for the cosmological energy shift.
For those GRBs showing consistent measures of $V_{f=0.45}$
in all channels we considered the weighted average.

\section{Results}
\label{s:results}
The selection of the GRBs with a significant and consistent
measure of variability reduced the sample from 827 to 551 BNB04 GRBs.

\begin{figure*}
\centerline{\includegraphics[width=14cm]{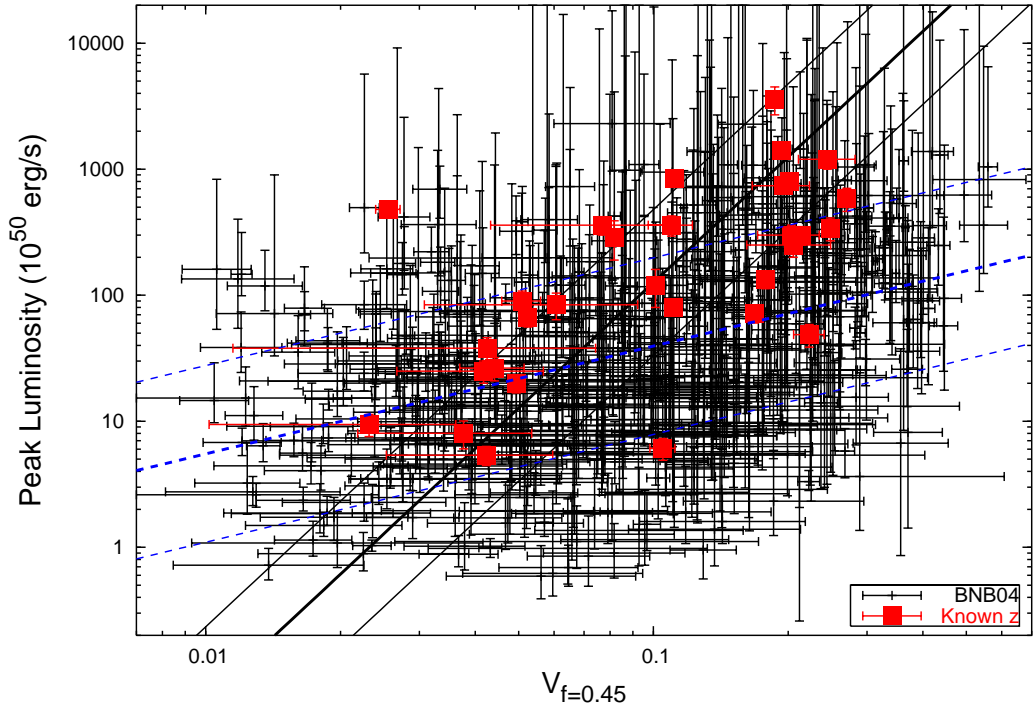}}
\caption{$V_{f=0.45}$ vs. Peak Luminosity for 551 BNB04 GRBs with significant
variability. Black solid lines mark the best-fitting power law found by
R01 (central line) and  $\pm 1\sigma$ widths; blue dashed lines show the result that we
found for 551 BNB04 GRBs. Red square points are GRBs with spectroscopic
redshift taken from G05. For scale compression reasons the subluminous GRB~980425 is
not shown.}
\label{f:alles_BNB04_wave_var_signif}
\end{figure*}

Figure~\ref{f:alles_BNB04_wave_var_signif} plots $V_{f=0.45}$
vs. $L$ for the subset of 551 BNB04 GRBs.
We also plot 31 GRBs with known redshift derived by G05
using data from GRBM/{\em BeppoSAX} \citep{Feroci97,Frontera97},
BATSE/{\em CGRO} \citep{Paciesas99}, FREGATE/{\em HETE-II} \citep{Atteia03},
{\em Ulysses} \citep{Hurley92}, Konus/{\em WIND} \citep{Aptekar95}
and BAT/{\em Swift} \citep{Gehrels04}. 
From the sample of 32 GRBs studied by G05 we ignored the peculiar subluminous
GRB~980425, not shown in Fig.~\ref{f:alles_BNB04_wave_var_signif}
for scale compression reasons.
%
%
The correlation between $V$ and $L$ is confirmed with higher significance due
to the high number of GRBs: the linear correlation coefficient is $0.437$
(P-value of $5\times 10^{-27}$), the Spearman's coefficient is $0.449$
(P-value of $10^{-28}$) and the Kendall's coefficient is $0.302$
(P-value of $2\times 10^{-26}$).

If we try to fit the correlation with a power law (eq.~\ref{eq:PL}),
the result is unsatisfactory ($\chi^2$/dof$=4238/549$).
\begin{equation}
\displaystyle \log{L_{50}} \ = \ m\,\log{(V_{f=0.45})} \ + \ q
\label{eq:PL}
\end{equation}
Similarly to G05 we conclude that a power-law fit
is inadequate to describe to correlation between $L$ and $V_{f=0.45}$.
Nevertheless, if one fits it with a power-law model,
the best-fitting index turns out to be $m=0.85\pm0.02$.
This is in contrast with the value originally found by R01 for a sample of 13 GRBs
with known redshift, $m_{\rm R01}=3.3^{+1.1}_{-0.9}$, and consistent
with that found by G05 for a sample of 32 GRBs with known
redshift, $m_{\rm G05}=1.30^{+0.84}_{-0.44}$.
Thus we infer that the correlation between $L$ and $V_{f=0.45}$ appears to
be shallower than that found by R01 and this confirms
what G05 found for the GRBs with known redshift.

\begin{figure*}
\centerline{\includegraphics[width=14cm]{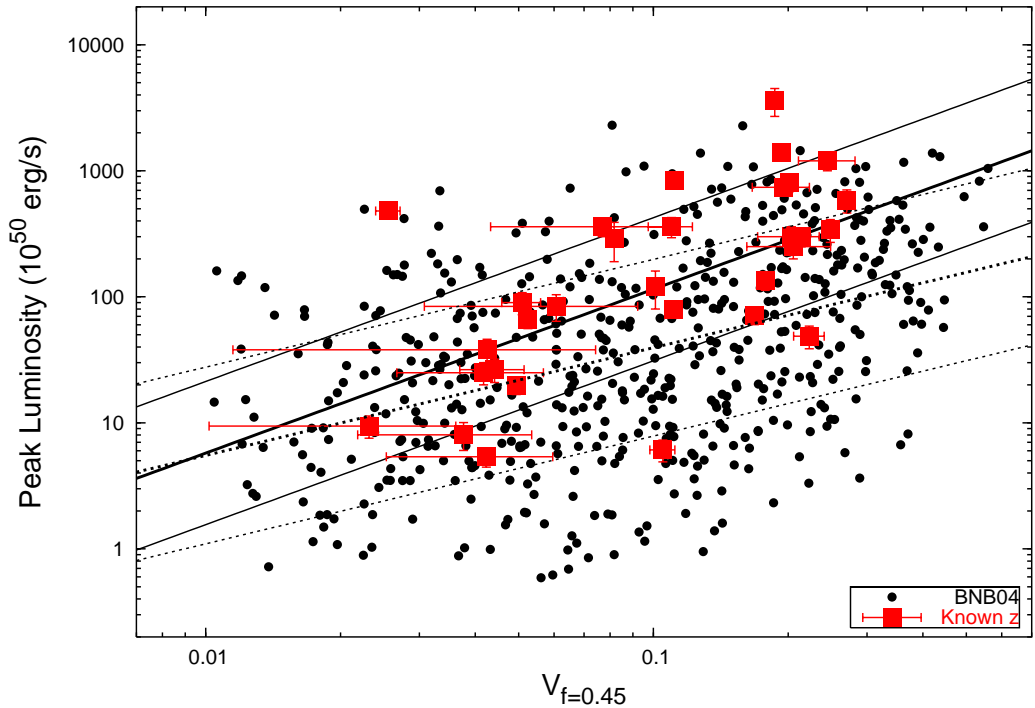}}
\caption{$V_{f=0.45}$ vs. Peak Luminosity for 551 BNB04 GRBs with significant
variability. This Figure shows the same data as Fig.~\ref{f:alles_BNB04_wave_var_signif},
aside from the error bars, here omitted for the sake of clarity.
Black solid lines mark the best-fitting power law found by
G05 (central line) and  $\pm 1\sigma$ widths for a sample of 32 GRBs
with known redshift (red square points). For scale compression reasons the subluminous
GRB~980425 is not shown. Dashed lines show the result that we
found for 551 BNB04 GRBs.}
\label{f:nurpoints_alles_BNB04_wave_var_signif}
\end{figure*}

\begin{figure}
\centerline{\includegraphics[width=8cm]{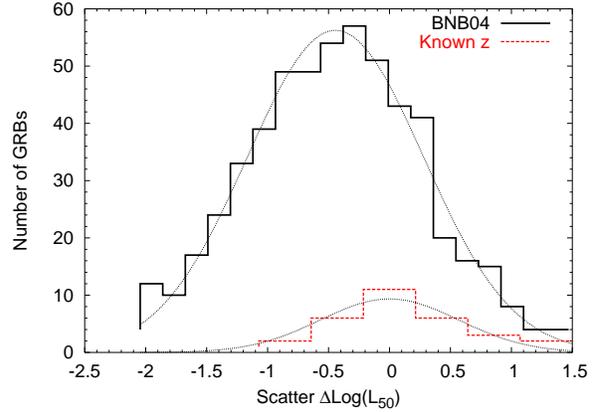}}
\caption{Distribution of the scatter of the two groups around the best-fitting
power law obtained by G05: 551 BNB04 GRBs (black solid line)
and 31 GRBs with measured redshift (red dashed line; the peculiar subluminous
GRB~980425 is not considered here). Also shown are the
two best-fitting normal distributions, whose $\sigma$'s are 0.70 (BNB04)
and 0.57 (G05).}
\label{f:scatter_wave_var_signif_both}
\end{figure}

Apparently these 551 BNB04 GRBs seem to locate differently from the 31 GRBs with
measured redshift (Fig.~\ref{f:nurpoints_alles_BNB04_wave_var_signif}).
This is mildly suggested by the result of a K--S test,
which gives $D=0.328$ with a probability of $9\times10^{-3}$ that the two groups
belong to the same class.
%
%
If we study the scatter distribution of the 551 BNB04 GRBs with respect to
the best-fitting power-law found by G05 for the GRBs with
known redshift, with $m_{\rm G05}=1.30^{+0.84}_{-0.44}$ and
$q_{\rm G05}=3.36^{+0.89}_{-0.43}$,
it comes out that the mean residual is -0.443
(in Fig.~\ref{f:nurpoints_alles_BNB04_wave_var_signif} this corresponds
to the residuals of the black circular points with respect to the black solid line).
Hence these GRBs are on average $10^{-0.443}\sim0.4$ times as luminous
as the sample of 31 GRBs with measured redshift.
The scatter of the 551 BNB04 GRBs is $\sigma=0.7$ to be compared with that
of the 31 GRBs with known redshift, which is $0.6$. The two distributions
are shown in Fig.~\ref{f:scatter_wave_var_signif_both}: the black solid line
shows the scatter distribution for the 551 BNB04 GRBs, while the red dashed
line shows the case of the 31 GRBs with measured redshift.
%
%

The results of the best-fitting power law and the scatter for the two groups
of GRBs (BNB04 and G05) are reported in Table~\ref{tab:fit_results}.
%
%
\begin{table}
\centering
  \caption{Best-fitting parameters for the two GRB sets for which
we find significant correlation between variability and peak luminosity. $m$ and $q$
are the parameters of the best-fitting power-law: $L_{50} = 10^q\,V_{f=0.45}^m$.
The last column reports the scatter around the best-fitting power law.}
  \label{tab:fit_results}
  \begin{tabular}{llllll}
\hline
GRB Set       & $m$ & $q$ & $\chi^2$/dof & scatter\\
              &     &     &              &        \\
\hline
551 BNB04     & $0.85\pm0.02$ & $2.45\pm0.03$ & 4238/549 & 0.70\\
 31 G05       & $1.30^{+0.84}_{-0.44}$ & $3.36^{+0.89}_{-0.43}$ & 1167/30 & 0.57\\
\hline
\end{tabular}
\end{table}

\section{Discussion}
\label{s:disc}
In agreement with the results by G05, the correlation between
variability and burst-frame 100--1000~keV peak luminosity originally discovered
by R01 is confirmed also using a sample of 551 BNB04 GRBs with a significant and
consistent measure of variability, as defined by R01.
Like in the case of the GRBs with known redshift considered by G05,
we find that the best-fitting power-law slope (index of $m=0.85\pm 0.02$) is
shallower than that found by R01 ($m_{\rm{\small R01}}=3.3_{-0.9}^{+1.1}$).
However, like G05 we find that the power-law description
of this correlation is inadequate ($\chi^2$/dof of 4238/549).
The result obtained for 551 GRBs with pseudo-redshift derived from the lag/peak
luminosity relationship is in agreement with that found by G05
for a sample of 32 GRBs with spectroscopic redshift and is incompatible
with the best-fitting power law obtained by R01.

The comparison between the sample of 551 BNB04 GRBs and that of the
GRBs with measured redshift considered by G05 reveals that the former
group is on average $0.4$ times as luminous as the latter.
Interestingly, despite the fact that a power law
does not provide a good description of the correlation, the two groups
show compatible power-law indices: $0.85\pm 0.02$ (BNB04) vs.
$1.30^{+0.84}_{-0.44}$ (G05).

We conclude that the disagreement found by G05 on a sample of 32 GRBs with known
redshift with respect to the original results by R01 derived from a sample of
13 GRBs with known redshift is confirmed by the sample of 551 GRBs
with pseudo-redshifts estimated from the spectral lag/peak luminosity
anticorrelation.

A thorough discussion and possible explanations of the discrepancy between
the results obtained by R01 and those by G05, consistent with those presented in this
paper, is reported in G05. However, we remark that the results derived
by R01 were based on a sample of only 13 GRBs with measured redshift available
at the time, some of which have just limits on the peak luminosity.
Remarkably, despite the fact that the sample of 32 GRBs with known redshift studied
by G05 collects light curves from different spacecraft and the sample here considered
of 551 BNB04 GRBs consists of BATSE data only, the results are consistent
with each other.

The fact that the BNB04 GRBs are on average less luminous than the 31 GRBs
with measured redshift could reflect the following possible observational bias:
the more luminous GRBs are more likely to have measurable redshift at lower wavelengths.

In summary, the two samples of GRBs show the same properties in
the variability/peak luminosity space: in either case the correlation
is confirmed, although it turns out to be inconsistent with that found
by R01. The only difference, aside from the lower average luminosity
of the BNB04 sample already discussed, is a little higher scatter of the
BNB04 GRBs around the best-fitting power-law: $\sigma=0.7$ to be compared
with $\sigma=0.6$ of the 31 G05 GRBs.

It is also worth mentioning that, when we selected the 551 BNB04 GRBs out of a sample
of 827 GRBs by requiring significant and consistent measures of variability
across different energy channels for each single GRB, a considerable fraction
of the 827 GRBs do not match these criteria. Thus, unlike the findings
by R01, we find that there are GRBs whose variability as defined by R01
does depend on the energy channel.

These results are not necessarily in contradiction with those by \citet{Schaefer01}
derived from a sample of BATSE GRBs. First of all, because the definitions
of variability and peak luminosity adopted by those authors are taken from FRR00.
Regarding the definition of variability, from fig.~3 of FRR00
we can estimate a power-law relation between the two measures of variability
according to the following: $V_{\rm FRR} \propto V_{\rm R01}^{\delta}$,
with $\delta\sim0.8-0.9$. The best-fitting power law found by \citet{Schaefer01}
between $L_{\rm FRR}$ (peak luminosity as defined by FRR00) and $V_{\rm FRR}$
has an index of $m_{\rm S01}=2.5\pm1.0$. Then it is:
$L_{\rm FRR}\propto V_{\rm R01}^{\delta\,m_{\rm S01}}\sim
V_{\rm R01}^{2.1\pm0.9}$. It must be pointed out that the relation between
the two definitions of variability should be investigated in detail, since
the above approximation is based on fig.~3 of FRR00, based on 8 GRBs only.
Nonetheless, this shows that the power-law index between variability and peak
luminosity does depend on the definitions adopted.
Moreover, the definition of variability given by FRR00 appears to be disputable
in some points: e.g., the action of rebinning by a non-integer factor the light curves
to report them to a fixed reference frame, dramatically affects the nature of Poisson
counting statistics of the time series. Consequently, the variance that must be
subtracted in the variability expression (eq.~2 in FRR00) is no more Poissonian
and should be corrected accordingly. In general, the operation of splitting the
counts integrated over a single time bin and use them to evaluate the
variability is potentially risky, especially when the smoothing time-scale,
proportional to the GRB duration, is relatively short.

\section{Conclusions}
We derived a selected sample of 551 BATSE GRBs from the BNB04 catalogue of GRBs
assuming pseudo-redshifts based on the peak luminosity/spectral lag anticorrelation
\citep{Norris00}.
The GRBs of this sample have been selected out of a sample of 827 BNB04 GRBs
by requiring a significant and consistent measure of variability across the
different energy channels, as defined by R01.
Unlike R01, we find that not all of the 827 BNB04 GRBs we initially considered
show a consistent measure of variability for different energy bands.

We confirm the correlation between variability and peak luminosity for
the subsample of 551 BNB04 GRBs. In agreement with the results by G05
on a sample of 32 GRBs with measured redshift and in contrast with the
original results by R01 on a sample of 13 GRBs with known redshift, we 
find that a power-law description of the correlation is inadequate.
Nonetheless, if we try to fit it, we obtain
a power law which is significantly shallower ($m=0.85\pm0.02$) than
that found by R01 ($m=3.3_{-0.9}^{+1.1}$) and consistent with that
found by G05 ($1.30^{+0.84}_{-0.44}$).

Finally, we note that the sample of 551 BNB04 GRBs are on average less
luminous by a factor of $\sim0.4$ with respect to the GRBs with
measured redshift. We ascribe this difference to the fact that the sample
of GRBs with measured redshift has been biased so far in favour of the more
gamma-ray luminous GRBs, which on average had high fluences and then
for which, consequently, more precise localisations were possible.
This is probably true for the very first GRBs with measured redshift.
We expect that {\em Swift} will clarify this issue as soon as it will
discover a number of GRBs with low peak luminosity and measurable
redshift.

\section*{Acknowledgments}
The author wishes to thank the anonymous referee, whose comments
contributed significantly to improve the quality of the paper and make it more
readable.
The author acknowledges his Marie Curie Fellowship from the European Commission.
This research has made use of data obtained
from the High-Energy Astrophysics Science Archive Research Center (HEASARC),
provided by NASA Goddard Space Flight Center.


\bsp

\label{lastpage}

\end{document}